\documentclass[letterpaper]{article}
\usepackage{epsfig}
\usepackage{graphicx}
\usepackage{url}
\usepackage{aaai-arXiv}
\usepackage{times}
\usepackage{helvet}
\usepackage{courier}
\usepackage{color}

\newcommand{\ignore}[1]{}

\begin{document}

\title{Identifying Duplicate and Contradictory Information in Wikipedia}
\author{
Sarah Weissman, Samet Ayhan, Joshua Bradley, and Jimmy Lin \\[1ex]
University of Maryland\\
College Park, Maryland, USA
}
\maketitle
\begin{abstract}
\begin{quote}
Our study identifies sentences 
in Wikipedia articles that are either identical
or highly similar by applying techniques for near-duplicate
detection of web pages. This is accomplished with a MapReduce
implementation of minhash
to identify clusters of sentences with high Jaccard similarity. We
show that these clusters can be categorized into six different types, two
of which are particularly interesting:\ identical sentences
quantify the extent to which content in Wikipedia is copied and
pasted, and near-duplicate sentences that state contradictory facts
point to quality issues in Wikipedia.
\end{quote}
\end{abstract}

\section{Introduction}

Readers of Wikipedia often notice that multiple articles contain
highly-similar or even identical passages. In some cases these represent
duplicate articles marked for merging, but content overlap arises
in other cases as well. 
For example, the article about a hurricane and the
article about the location where it made landfall might share the same
content about the impact of the natural disaster. Identical content is
most likely the result of copy and paste between articles, but
interestingly, readers occasionally come across highly-similar content that
state contradictory facts. In a distributed environment where anyone can
edit content, these observations are perhaps not surprising, 
and in this paper we attempt to
rigorously characterize these phenomena by treating the
problem as that of near-duplicate sentence detection. We
adapt standard locality-sensitive hashing (LSH) techniques to identify
clusters of near-duplicate sentences in Wikipedia.

We believe this problem is interesting in a few ways:\ 
For duplicate sentences,
our analyses quantify the extent to which Wikipedia content is simply
{\it replicated}, as opposed to {\it written} from scratch. 
In the case of {\it near} duplicates, some differences represent
minor copyediting that does not change the substance of the content, but
in other cases the differences 
represent contradictory facts. Quantifying these cases
provides an indirect measure of the quality of Wikipedia in terms of
self consistency. 

This work makes no claims about the novelty of our
techniques nor our implementation in MapReduce.
Rather, our contribution lies in the analysis and categorization
of near-duplicate sentence types.
We have not seen
locality-sensitive hashing applied to Wikipedia in this way before.

\section{Related Work}

The problem we tackle in this paper is related to a few other problems
that have been studied before. Near-duplicate 
detection of web pages~\cite{Henzinger_SIGIR2006} is important
in search because web pages are often copied or mirrored with only
minor differences (e.g., ads or navigation bars); it would be
desirable to return only the ``canonical'' versions in search
results. In fact, the algorithm that we use in this paper,
minhash~\cite{broder:resemblance}, was originally developed for
exactly this purpose. Another closely-related problem
is plagiarism detection~\cite{Si:1997:CDP:331697.335176}, 
or more generally, ``text
reuse''~\cite{bendersky:timeline}. In contrast to near-duplicate detection, the
focus is usually on smaller segments of text as opposed to entire
documents. However, similar approaches such as shingling are
applicable to both problems.

Other similar formulations of the problem are what the data mining
community calls pairwise similarity search or ``all pairs''
search~\cite{Bayardo_etal_WWW2007} and what the database community
calls set similarity join~\cite{Vernica_etal_SIGMOD2010}. The task is the same:\ given a
potentially large collection of objects, identify all pairs whose
similarity is above a threshold according to some similarity metric.

There are two classes of solutions to the above problems:\ in
the {\it index-based} approach, an inverted index is constructed from
objects in the collection and a traversal of the index allows the
similar pairs to be extracted,
e.g.,~\cite{Bayardo_etal_WWW2007,lin:brute}; with the {\it hash-based}
approach, the basic idea is to use locality-sensitive hashing (LSH) to
identify similar pairs based on hash collisions, e.g., 
minhash~\cite{broder:resemblance}. Of course, hybrid solutions are also
possible. Scaling up these solutions has been
accomplished by MapReduce~\cite{lin:brute,Vernica_etal_SIGMOD2010}.
Similarly, our approach takes advantage of minhash using a MapReduce
implementation in Hadoop.

Because of its open nature, Wikipedia has generated
much controversy over its editorial quality and factual correctness. 
An early study found Wikipedia's accuracy to rival that of
traditional encyclopedias~\cite{Giles:wiki}, but subsequent investigations
have arrived at conflicting conclusions. A thorough review
is beyond the scope of this short paper, but somewhat ironically,
the best summary of this ongoing debate is a Wikipedia article.\footnote{
  http://en.wikipedia.org/wiki/Reliability\_of\_Wikipedia}
Since Wikipedia may be edited anonymously,
information may be freely copied between web sources and even between
Wikipedia articles without verification---however, this is not to
say that there are no quality assurance mechanisms in Wikipedia~\cite{Stvilia:2008}.
Although there are active communities of
editors who contribute to the upkeep of various articles, much
of Wikipedia is edited and expanded in an ad hoc manner.
In particular,
Wilkinson and Huberman~\shortcite{wilkinson:wiki} 
found the distribution of article edits
on Wikipedia to have a long tail, meaning that a small number of
articles account for most of the edits, and that the
number of edits is related to article quality. Articles with
few edits and low editorial attention are less likely to be updated,
which is a source of contradictory information based on our analysis.

\section{Near-Duplicate Sentence Detection}

For near-duplicate detection we use a well-known technique called
minhash~\cite{broder:resemblance}. We begin with a parameterized family
of $N$ hash functions $F_i$, $1 \le i \le N$. Each sentence in a
Wikipedia article is broken up into $n$-gram ``shingles'' (at the
character level) and for the shingle set $S$ a set $\{min_{s \in
  S}(F_i(s)\}$ of minimum hashes over the hash family is generated.
The signature of a document $d$ is represented as a vector of $K$
minhashes randomly selected from the set of $N$. To increase recall we
generate $M$ signatures for each sentence (i.e., $M$ draws of $K$ from
$N$). Broder proves a straightforward relationship between minhash
collisions (i.e., documents that share the same signature)
and their Jaccard similarities, which forms the theoretical foundation
of why and how minhash ``works''; we refer the reader to the
original paper for the relevant proofs.

\subsection{MapReduce Implementation}

We implemented minhash in MapReduce~\cite{Dean_Ghemawat_OSDI2004} using Hadoop
for this study.
The algorithm is as follows:\
each mapper receives a Wikipedia article
identified by a unique docid. Inside the mapper
we break the article into sentences using a regular expression;
sentences that are shorter than 75 shingles or longer than 600
shingles are discarded. For each sentence, the $M$ minhash signatures are
then computed (per above). The family of hash functions is implemented using a
``Multiply Shift'' hashing scheme\footnote{\small
  http://en.wikipedia.org/wiki/Universal\_hashing} and generated
from a random seed. 
For our experiments we
use a 60 bit hash and a hash family of size $N=20$.
Each signature is emitted as the key of an intermediate key--value
pair with the sentence id as the value (constructed from the docid and
the sentence number).

The MapReduce programming model guarantees that all values associated
with the same key (minhash signatures in our case)
are shuffled to the same reducer and grouped together for further
processing---in effect,
collecting the hash collisions for us. In the reducers we receive
signatures as keys and all sentence ids that share a
signature as values. If there is more than one value per key, we write out all
sentence ids as a cluster. This serves as input to the final cluster
generation stage (more below).

\subsection{Parameter Tuning}

\begin{figure}
\centering
\includegraphics[width=2.75in, keepaspectratio = true]{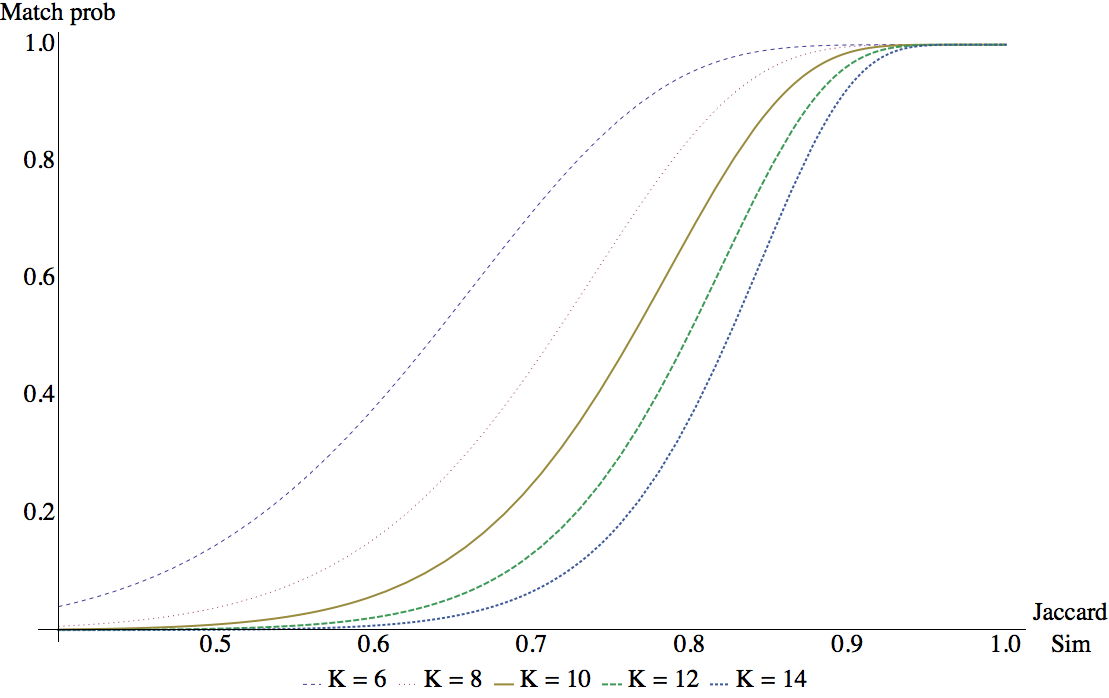}
\vspace{-0.25cm}
\caption{Minhash Parameter tuning.}
\label{clust}
\vspace{-0.3cm}
\end{figure}

One complexity of applying minhash to real-world datasets is the
myriad of parameters that must be selected---each setting manifests a
tradeoff between precision, recall, and computational effort.
Our approach to parameter tuning relied on a
combination of analytical calculations and hand-tuning based on examining
the output. We began by fixing the hashing scheme and the size of the
hash family ($N=20$). We then generated
10 signatures ($M$ = 10) per input sentence.
Based on Broder~\shortcite{broder:resemblance}, the
probability of a match for sentences $A$
and $B$ can be expressed as follows:
\[P[match(A,B)] = 1 - (1 - s^K)^M\]
where $s = \textrm{Jaccard}(A,B)$.
With the above settings, the effects of different $K$'s are shown
in Figure~\ref{clust}. Based on this analysis, we set $K=10$.
This means that if we choose 0.9 Jaccard similarity as our goal (90\% overlap
in shingle sets), then there is a 99\% chance of a match (i.e., the recall). 
Finally, after some hand tuning, we settled on a shingle length of 12
characters. This setting means that we obtain
shingles that cross word boundaries, which allows us to capture (to some extent) word
order in our similarity computations.

Based on the parameter settings, it is possible to estimate the amount
of data that is generated by our approach, which is the product of the
length of each signature, the number of signatures per sentence, and
the total number of sentences. This computation is useful because the
amount of intermediate data provides a rough proxy for algorithm
running time.

Finally, note that it is possible to improve precision by filtering
results with a second pass on the output signature groups. In this
second pass we can discard false positives or apply another
similarity metric (e.g., edit distance). Compared to the computational
cost of minhash, such additional processing is
cheap since it is applied to far less data. However, we
did not implement second-pass filtering in our experiments
and leave this for future work.

\subsection{Final Cluster Generation}

The output of minhash is a set of clusters, where each
cluster represents a signature collision. Since we generate multiple
signatures per sentence, it is possible that
a sentence appears in multiple clusters. We adopt the standard
practice of merging all clusters that share at least one common
sentence.
Cluster merging is accomplished in one pass 
(outside MapReduce) with a union find data
structure~\cite{clr}. 
In union find each node maintains a pointer to a head node
for the set. We maintain a lookup table from each sentence id to its
node in the union find structure. Iterating over the input clusters, we
look up the node for each sentence, merging clusters or creating new
ones as needed. Merging two clusters involves changing the head node
of one cluster to point to the head of the other. 
Once all clusters are processed
we perform a pass over the lookup table to obtain the mapping from
clusters to nodes. Another MapReduce job reconstructs the
sentences in each cluster.

\begin{figure}
\centering
\includegraphics[width=3.25in, keepaspectratio = true]{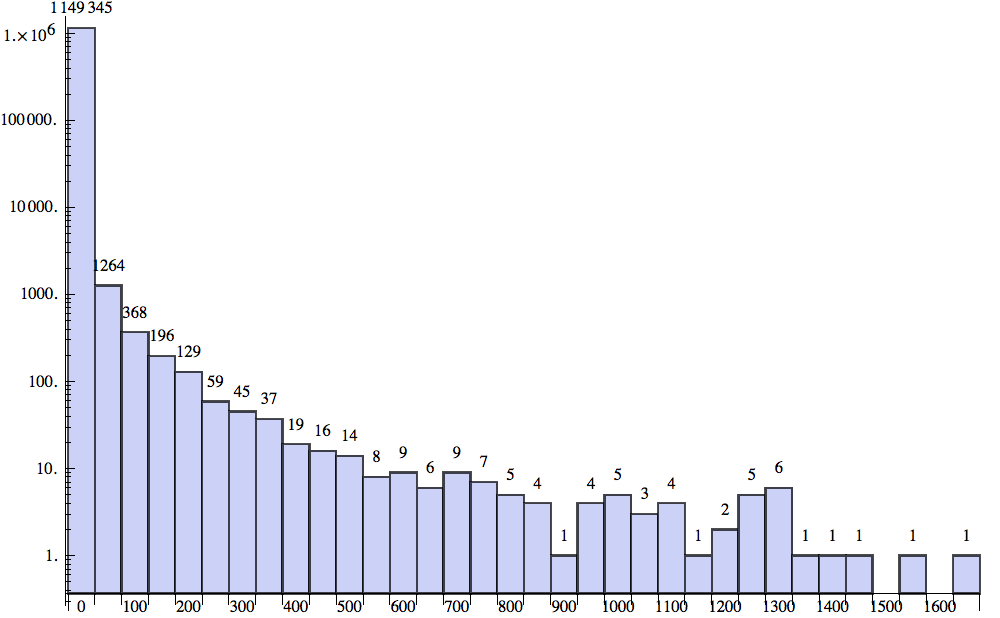}
\vspace{-0.2cm}
\caption{Histogram of cluster sizes. Bars show bucketed cluster sizes and
heights show number of clusters of that size (log scale).
The 25 clusters of size $\ge 1700$ are not shown.}
\vspace{-0.2cm}
\label{params}
\end{figure}

\section{Experimental Results}

Our analysis was conducted on an XML dump of English Wikipedia from
July 2013. The entire corpus is approximately 42 GB and contains 10.2m
articles (after excluding certain non-article pages). Based on our
simple sentence chunker, there are a total 135.8m sentences.

Running our Hadoop implementation of minhash on the entire collection
took approximately 25 minutes on our cluster consisting of 16 nodes (128
cores). Code necessary for replicating these experiments have been
open sourced.\footnote{https://github.com/seweissman/wikiduper}

After the processing pipeline, we identified 1.15m clusters, 3.50m
article/sentence pairs over 1.09m articles. The clusters contain 2.36m
unique sentences. Cluster sizes range from 2 to over 40k; see
Figure~\ref{clust} for a histogram. Most clusters are small; about 99\%
of the clusters have ten or fewer sentences, but about a
quarter of the output sentence/article pairs fall into clusters with
sizes greater than ten. 

\begin{figure*}[t]
{\small
{\bf Templates}\\
Of the agricultural land 40.4\% is used for growing crops and 26.6\% is pastures while 2.2\% is used for orchards or vine crops. [Gondiswil] \\
Of the agricultural land 26.1\% is used for growing crops and 30.2\% is pastures while 3.0\% is used for orchards or vine crops. [Kleindietwil] \\
Of the agricultural land 37.8\% is used for growing crops and 35.5\% is pastures while 2.2\% is used for orchards or vine crops. [Leimiswil] \\[1ex]
{\bf Identical}\\
Professional organizers help redirect paradigms into more useful cross-applications that ensure properly co-sustainable futures for their clients' spaces and processes. [Professional organizing] \\
Professional organizers help redirect paradigms into more useful cross-applications that ensure properly co-sustainable futures for their clients' spaces and processes. [Professional organizer] \\[1ex]
{\bf Copyediting}\\
In dry areas it may only emerge from its burrow for a few weeks when conditions are right and usually at night but in areas with permanent water bodies and abundant rain it may be active all day. [Great Plains toad] \\
In dry areas it may only emerge from its burrow for a few weeks when conditions are right and only at night but in areas with permanent water bodies and abundant rain it may be active all day. [List of amphibians and reptiles of Montana]\\[1ex]
{\bf Factual Drift}\\
Bulgaria, a poor rural nation of 7 million people sought to acquire Macedonia but when it tried it was defeated in 1913 in the Second Balkan War. [History of the Balkans] \\
Bulgaria a poor rural nation of 4.5 million people sought to acquire Macedonia but when it tried it was defeated in 1913 in the Second Balkan War. [Home front during World War I] \\[1ex]
{\bf References}\\
Komj\'{a}th P\'{e}ter and Vilmos Totik: Problems and Theorems in Classical Set Theory, Springer-Verlag, Berlin, 2006. [P\'{e}ter Komj\'{a}th] \\
P\'{e}ter Komj\'{a}th, Vilmos Totik: Problems and Theorems in Classical Set Theory, Springer-Verlag, Berlin, 2006. [Vilmos Totik] \\[1ex]
{\bf Other}\\
Army Medical Research Institute of Infectious Diseases (USAMRIID) microbiologist Bruce E. [Timeline of the 2001 anthrax attacks] \\
Army Medical Research Institute of Infectious Diseases (USAMRIID) which transitioned from the previous U.S. [Fort Detrick]}
\caption{Classification of near-duplicate sentences in Wikipedia, with examples (article titles in brackets).}
\label{toparticlegroups}
\end{figure*}

Based on manual inspection of the cluster output, we can categorize
near-duplicate clusters into one of six types. Examples are shown in
Figure~\ref{toparticlegroups}
and discussed below.

\emph{Templates} describe sentences that have identical structure, but
with different entities, facts, or figures for {\it different} topics
(and thus are not contradictory). They reflect conscious attempts to
impose structure across groups of articles that may be related. Since
several of the largest template clusters contain tens of thousands of
sentences (the largest over 40,000), it is likely that some
template groups are automatically generated using bots. In many cases
template sentences are found in stub articles. 

\emph{Identical} sentences are the result of copy and paste, and are
often found in articles that cover similar topics or articles that are
subtopics of other topics.
Non-identical but highly-similar sentences break down into two types:
\emph{Copyediting} refers to nearly identical sentences that differ in
stylistic or otherwise non-substantive ways.
They arise with minor
editing after a copy and paste. \emph{Factual drift} describes
sentences about the {\it same} topic that provide contradictory
facts. Although without detailed research, there is no way to
ascertain which version (if any) is correct, we can identify a common
scenario. After a copy and paste, a fact becomes out of date (e.g., the
tallest building or the death toll in a disaster) and is corrected in
one instance but not the others.

\emph{References} refer to citations, typically occurring at the end
of articles. Since Wikipedia does not adhere to one single citation
style, the same work may be cited differently, or multiple citations
to the same venue may be similar.

Finally, clusters that do not fit into any of the above categories
are classified as \emph{other}. These typically represent sentences
that are highly similar, but otherwise bear no semantic relationship
with each other. Sentence chunking errors often contribute to these
spurious results.

To quantify the distribution of these six cases, we randomly sampled
2094 clusters and performed manual classification. The results are
shown in Table~\ref{counts}. Nearly three quarters of all clusters
are either \emph{identical} or \emph{templates}.
From this table, we obtain a rough characterization of the precision
of our analysis based on minhash---except for ``other'', all categories are
``correct'' identification of near-duplicate sentences. Recall can
be computed analytically, as we have shown in the previous section.

\begin{table}
\centering
\caption{Manual classification of sample clusters.}
\vspace{0.2cm}
\begin{tabular}{| l | r | r |}
\hline
type & count & fraction \\
\hline
\hline
Templates     & 632 & 30.2\% \\ \hline
Identical     & 948 & 45.3\% \\ \hline
Copyediting   & 283 & 13.5\% \\ \hline
Factual drift & 121 &  5.8\% \\ \hline
References    &   7 &  0.3\% \\ \hline      
Other         & 103 &  2.9\% \\ \hline
\end{tabular}
\label{counts}
\vspace{-0.2cm}
\end{table}

Finally, we manually examined a few of the large clusters and noticed
that they often contain a mix of different phenomena. One common
pattern is distinct groups of identical sentences, where each of the
groups are near duplicates. Since there are relatively few large
clusters, these nuances do not have a significant impact on the
figures in Table~\ref{counts}.

\section{Future Work and Conclusions}

In this work, we applied minhash to the
problem of near-duplicate sentence detection on Wikipedia. Our
MapReduce implementation is highly scalable and processes English
Wikipedia in a short amount of time on a modest cluster. We found that
there is a substantial amount of duplicate content in Wikipedia, and
that near-duplicate sentences manifest a few phenomena, the most
interesting of which is contradictory facts. How can we act upon this
analysis? We could imagine a robot that monitors Wikipedia to flag
inconsistencies and requests editors to intervene and
resolve. Such a service would be valuable in improving the
internal consistency and quality of Wikipedia.

There are several directions in which our work can be extended.
Currently, our technique only identifies \emph{clusters} of near-duplicate
sentences---missing from our
analysis is the notion of information flow:\ Which was the source
article and which was the target of copying and pasting? 
Are there copy ``chains'' where
content was progressively copied from one article to the next, with
possible ``branches'' (and accumulation of errors along the way)? 
Our analysis of the large 
near-duplicate clusters suggests that there are complex 
edit histories that form tree-like structures.
Furthermore, are there editor-specific effects? For example,
is copying and pasting more likely by anonymous editors?

Finally, templates represent interesting cases of near-duplicate sentences
because they, in effect, encode structured knowledge.
We could formalized such knowledge in infoboxes if they are not
already:\ it might be possible to apply information extraction
techniques to generate structured linked data from template sentences. Going in
the opposite direction, we might (via a robot) {\it create} templates
sentences to impose consistency across pages in the same category to ensure
that certain facts are captured in prose.

\bibliographystyle{aaai}
\small

\end{document}